\documentclass[10pt,dvips,fleqn]{article}

\usepackage{amsmath}
\usepackage{amsfonts}
\usepackage{amssymb}

\usepackage[dvips]{graphicx,color}

\usepackage[hang,footnotesize]{caption2}

\usepackage[colorlinks,dvips,citecolor=blue,linkcolor=blue,urlcolor=blue]{hyperref}

\newcounter{propCounter}
\newenvironment{prop}[1][Proposition]{\refstepcounter{propCounter} \vspace{1.5ex} \noindent \textbf{#1 \arabic{propCounter}.}
\itshape}{\vspace{1ex}}
\newenvironment{proof}[1][Proof]{\noindent \textbf{#1:}
}{ \hfill \rule{0.5em}{0.5em} }
\newenvironment{remark}[1][Remark]{\vspace{1ex} \noindent \textbf{#1:} \slshape}{\vspace{1ex}}

\setcaptionwidth{.8\textwidth}

\newcommand{\R}{\ensuremath{\mathbb{R}}}
\newcommand{\F}{\ensuremath{\mathcal{F}}}

\newcommand{\Sig}{\ensuremath{\Sigma}}
\newcommand{\di}{\ensuremath{\partial}}

\newcommand{\ti}{\ensuremath{\tau_{i}}}
\newcommand{\tf}{\ensuremath{\tau_{f}}}

\begin{document}

\DeclareGraphicsExtensions{.eps}

\title{Bohmian transmission and reflection dwell times without trajectory sampling}

\author{Sabine Kreidl\\[10pt]{\normalsize Institut f\"ur Theoretische Physik, Universit\"at Innsbruck}\\
{\normalsize Technikerstr. 25, A-6020 Innsbruck,
Austria}\\[2pt]{\normalsize \href{mailto:sabine.kreidl@uibk.ac.at}{sabine.kreidl@uibk.ac.at}}}

\date{}

\maketitle

\begin{abstract} \noindent Within the framework of Bohmian mechanics dwell times find a
straightforward formulation. The computation of associated
probabilities and distributions however needs the explicit
knowledge of a relevant sample of trajectories and therefore
implies formidable numerical effort. Here a trajectory free
formulation for the average transmission and reflection dwell
times within static spatial intervals $[a,b]$ is given for
one-dimensional scattering problems. This formulation reduces the
computation time to less than 5\% of the computation time by means
of trajectory sampling.\\[15pt]PACS numbers: 03.65.-w
\end{abstract}

\section{Introduction}

\setcounter{footnote}{-1}

\begin{figure}[h!]
\centering
\includegraphics[width=.4\textwidth]{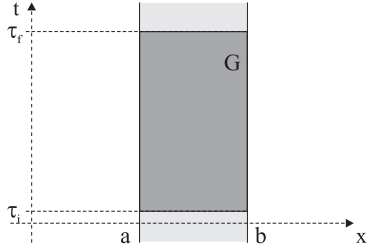}\\
\caption{Spacetime region $G = [a,b] \times
[\ti,\tf]$.\label{1dRID}}
\end{figure}

For a 1D static detector located in the spatial interval $[a,b]$
(see figure \ref{1dRID}), the {\it average dwell time} of an
ensemble of quantum systems with wave function $\Psi$ from time
$\ti$ up to time $\tf$ is generally agreed to be given by
\begin{equation}\label{1dDwellTime} \int_{\ti}^{\tf} dt
\int_{a}^{b} d\xi \; \left\vert \Psi(t,\xi) \right\vert^{2},
\end{equation}
cf the reviews \cite{Stovneng}\footnote[1]{Clearly expression
(\ref{1dDwellTime}) does not exist for bound states.}. It is
motivated by classical reasoning \cite{Muga3}, and it also has
been derived within Bohmian mechanics \cite{Leavens1}. In a recent
work \cite{Damborenea} a corresponding dwell time operator has
been investigated.

Dwell times of the type (\ref{1dDwellTime}) were associated with
interaction times in collision processes long ago \cite{Smith}.
The relevance of these interaction times to time resolved
scattering experiments has been studied, e.g., in
\cite{Nussenzveig}. Through these works it became clear, that
\emph{differences} of average dwell times formed between a
scattering wave packet and its free incoming asymptote are
measurable quantities.

With the introduction of the Larmor clock by Baz \cite{Baz1},
dwell time expressions themselves have got measurable status. The
idea is that a small and uniform magnetic field, which is confined
to a small region of space, causes a Larmor-precession of the
spin-polarization vector of the scattered wave. It was shown in
\cite{Martin} that the Larmor clock indeed reveals the average
dwell time. If it also is capable of displaying the spectral
distribution of some dwell time operator is still unclear.

The specialization of the Larmor clock to the case of
one-dimensional scattering was done by Rybachenko in \cite{Ryb}.
In this work (to the authors knowledge for the first time) a
distinction into the dwell times of the finally transmitted
respectively reflected partial waves has been introduced. For
short such selective dwell times will further on be denoted as
transmission and reflection times. Another approach to
transmission and reflection times, grounded on a specific
experimental scheme, was introduced by the oscillating barrier
model of B\"uttiker and Landauer \cite{Buettiker1}. This model is
widely believed to have ignited the tunnelling time controversy
anew. A further extension of the Larmor clock by B\"uttiker
\cite{Buettiker2} incorporates the effect of spin-alignment with
the magnetic field.

More recently the interest in transmission times has been driven
on the one hand by the very indirect time measurement techniques
of the condensed matter community, especially in connection with
tunnelling in semiconductor heterostructures or Josephson
junctions. On the other hand, the progress in laser cooling
techniques in quantum optics delivers another valuable tool for
future time resolved scattering experiments.

It is no surprise that the detailed definition of transmission and
reflection times depends on the situation under scrutiny. A
systematic operator approach embodying several such possible
definitions was given by Brouard \emph{et al} \cite{Brouard}. It
was shown, however, that transmission and reflection times derived
within the framework of Bohmian mechanics are not included in this
catalogue (\cite{Muga2}, chapter 5).

Bohmian mechanics comprises the mathematical concept of world
lines or trajectories and with this the term `particle' obtains
substance in quantum theory again. Thus the notion of dwell time
can be addressed in a straightforward manner, very much like in
classical mechanics. The dwell time of a particle in the spatial
interval $[a,b]$ is simply defined as the duration during which
the particle's trajectory is localized within $[a,b]$. For an
elaborate discussion of Bohmian mechanics see \cite{Duerr2}.

As the numerical effort involved with the calculation of Bohmian
world lines is immense, there have been attempts to compute the
Bohmian transmission and reflection times without the sampling of
trajectories. A related one-dimensional bound-state situation has,
e.g., been studied by Stomphorst in \cite{Stomphorst}. In the
present paper a formulation without trajectories for the
transmission and reflection times in a genuine scattering
situation is presented. The derivation closely follows ideas
developed for the treatment of 1D arrival time in \cite{Kreidl}.
As an application the scattering from a double potential barrier,
reminiscent of semiconductor-heterojunctions in, e.g., resonant
tunnelling diodes, is considered.


\section{Bohmian transmission and reflection times in 1D scattering situations  \label{sTRT}}

We consider a scattering situation in which the M\o ller operators
$\Omega^{in}$ and $\Omega^{out}$ exist and are asymptotically
complete \cite{ReedSimon}. $\Psi(t,\cdot)=e^{-iHt/ \hbar}
\Omega^{in} \phi_0= \Omega^{in} e^{-iH_0 t/ \hbar} \phi_0$ denotes
the scattering solution with incoming asymptote
$\phi(t,\cdot)=e^{-i H_0 t/ \hbar} \phi_0$. The solution
$\phi(t,\cdot)$ of the free Schr\"odinger equation is chosen to be
localized on the negative spatial semi-axis for $t \to -\infty$.
That is the case if and only if the Fourier transform
$\mathcal{F}(\phi_0)$ is localized on the positive half-line
\cite{Dollard}.

The Bohmian transmission time in the above scattering situation is
defined as follows. Let $\gamma_x(t)$ denote the Bohmian
trajectory which at time $t=0$ passes through the point $x$. There
exists a critical value $x_c \in \R$, such that $\lim\limits_{t
\to \infty} \gamma_x(t) = -\infty$ for all $x<x_c$ and that
$\lim\limits_{t \to \infty} \gamma_x(t) = \infty$ for all $x>x_c$.
Therefore the Bohmian transmission time is defined as
\begin{equation}\label{transT}
\langle \tau_T \rangle = \int\limits_{-\infty}^{\infty} dx \;
\vert \Psi(0,x) \vert^2 \cdot \Theta(x-x_c)
\int\limits_{\ti}^{\tf} dt \; \chi_{[a,b]}(\gamma_x(t))
\end{equation}
with $\chi_{[a,b]}$ the characteristic function on the interval
$[a,b]$ and $\Theta$ the Heaviside step function. See, e.g.,
\cite{Leavens2}. Analogously the reflection time $<\tau_R>$ is
defined by replacing the term $\Theta(x-x_c)$ in the right hand
side of equation (\ref{transT}) by $\Theta(x_c-x)$. The critical
trajectory $\gamma_{x_c}(t)$ ($\gamma_{x_c}(0)=x_c$) is implicitly
defined by
\begin{equation}\label{xc}
\vert T \vert^2 = \int_{\gamma_{x_c}(t)}^{\infty} \vert
\Psi(t,\xi) \vert^2 \; d\xi, \quad \forall t \in \R.
\end{equation}
Thereby
\begin{equation}\label{TransKoeff}
\vert T \vert^2 := \lim_{t \to \infty} \int_{0}^{\infty} \vert
\Psi(t,\xi) \vert^2 \; d\xi
\end{equation}
is the transmission probability of the scattering system. The
lower limit of the integral in (\ref{TransKoeff}) can equally be
replaced by any finite $q \in \R$. Accordingly by $\vert R
\vert^{2}:=1-\vert T \vert^2$, the reflection probability is
defined. Thus the conditional transmission respectively reflection
times, i.e. transmission and reflection times normalized to the
fraction of transmitted respectively reflected particles of the
entire ensemble, are $<\tau_T>^c:=\frac{1}{\vert T \vert^2}
<\tau_T>$ and $<\tau_R>^c:=\frac{1}{\vert R \vert^2} <\tau_R>$.

\noindent From the computational viewpoint, the two terms
$<\tau_T>$ and $<\tau_R>$ by means of Bohmian mechanics are
achieved in a straightforward manner. One chooses an appropriate
sample of initial values on the configuration space $\Sig_0=\R$,
calculates the corresponding trajectories over a sufficient range
of time, determines the dwell time for each trajectory, labels the
trajectories as transmitted or reflected according to their
position at large times and finally calculates, according to the
weight of each trajectory and just as in classical statistics, the
average times. That this programm involves formidable numerical
effort is evident.

However, the calculation of Bohmian transmission and reflection
times can be reduced to the computation of current density
integrals along the edges at $x=a$ and $x=b$. The next proposition
represents a generalization of expressions already proposed in
\cite{Oriols}.

\begin{prop} \label{propTRDT} For 1D
scattering solutions $\Omega^{in} \phi_t$ with $\Theta(K)
\phi_0=\phi_0$, $\Vert \phi_0 \Vert=1$, for the Bohmian
transmission and reflection times within $[a,b] \times [\ti,\tf]$,
hold
\begin{equation}\label{TransTime}
\langle \tau_T \rangle = \int_{\ti}^{\tf} dt \; \left[ \min\left\{
f_a(t), \vert T \vert^2 \right\} -  \min\left\{ f_b(t), \vert T
\vert^2 \right\} \right]
\end{equation}
and
\begin{equation}\label{ReflTime}
\langle \tau_R \rangle =  \int_{\ti}^{\tf} dt \; \left[
\max\left\{ f_a(t), \vert T \vert^2 \right\} -  \max\left\{
f_b(t), \vert T \vert^2 \right\} \right]
\end{equation}
with
\[
f_{q}(t):=\int_{-\infty}^{t} j(s,q) \; ds.
\]
Here $j$ is the quantum mechanical probability current density.
\end{prop}

\noindent An essential ingredient for the proof of proposition
\ref{propTRDT} is the relation
\begin{equation}\label{jrhoZshg}
f_{q}(t):=\int_{-\infty}^{t} j(s,q) \; ds = \int_{q}^{\infty}
\vert \Psi(t,\xi) \vert^2 \; d\xi.
\end{equation}
It depicts that the probability of finding a particle to the right
of $q$ at time $t$ is equal to the amount of probability, which
has passed $q$ up to the time $t$. A plausibility argument for
equation (\ref{jrhoZshg}) and a rigorous proof for a limited class
of scattering situations is given in appendix A. The proof of
proposition \ref{propTRDT} is given in appendix B.

\begin{remark}
The formulation of Oriols \emph{et al} \cite{Oriols} assumes the
case in which the current density at the right edge of the barrier
doesn't change its sign and is positive for all times. In this
case $f_b(t) \leq \vert T \vert^2$, $\forall t \in \R$, because
\[
\vert T \vert^2 \stackrel{(\ref{TransKoeff})}{=} \lim\limits_{t
\to \infty} \int\limits_{b}^{\infty} \vert \Psi(t,\xi) \vert^2 \;
d\xi \stackrel{(\ref{jrhoZshg})}{=} \lim\limits_{t \to \infty}
\int\limits_{-\infty}^{t} j(s,b) \; ds =
\int\limits_{-\infty}^{\infty} j(s,b) \; ds
\]
and further
\[
f_b(t)=\int\limits_{-\infty}^{t} j(s,b) \; ds \stackrel{j(\cdot,b)
\geq 0}{\leq} \int\limits_{-\infty}^{\infty} j(s,b) \; ds = \vert
T \vert^2.
\]
Then equation (\ref{TransTime}) reduces to
\[
\langle \tau_T \rangle^c = \frac{1}{\vert T \vert^2} <\tau_T> =
\frac{1}{\vert T \vert^2} \int_{\ti}^{\tf} dt \; \left[
\min\left\{ f_a(t), \vert T \vert^2 \right\} - f_b(t) \right]
\]
and equation (\ref{ReflTime}) to
\[
\langle \tau_R \rangle^c = \frac{1}{\vert R \vert^2} <\tau_R> =
\frac{1}{\vert R \vert^2} \int_{\ti}^{\tf} dt \; \left[
\max\left\{ f_a(t), \vert T \vert^2 \right\} -
 \vert T \vert^2 \right]
\]
which reproduces equations (14) and (15) of \cite{Oriols} for the
special choices $\ti=0$ and $\tf=\infty$.
\end{remark}

\noindent Obviously an interesting task would be to construct a
device, i.e. a clock, which measures the Bohmian transmission and
reflection times. Such a clock should display the respective time,
let us say, through the Bohmian center of mass position of its
hand at the instant of its readout, which presumably has to be
chosen by the experimenter. The combined system's wave function
would be modelled by an appropriate Schr\"odinger equation,
incorporating the interaction between the micro-system and the
clock. There is a self-adjoint operator corresponding to the
hand's positions. The crucial question is whether this
observable's spectral distribution in a given state at the instant
of its readout coincides with the respective Bohmian transmission
(reflection) time distribution. Probably this is not the case for
all states. However, similarly to the issue of exit time
statistics \cite{Duerr1}, a subspace might be identified on which
the Bohmian distribution coincides with one of the standard
quantum mechanical distributions.

\section{ Numerical example: transmission and reflection at a double potential
barrier structure}\label{sExamples}

\begin{figure}[h!]
\centering
\includegraphics[width=.4\textwidth]{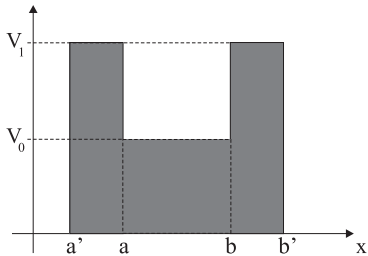}
\caption{Double potential barrier.\label{DBPotential}}
\end{figure}

\begin{figure}[h!]
\centering
\includegraphics[width=.8\textwidth]{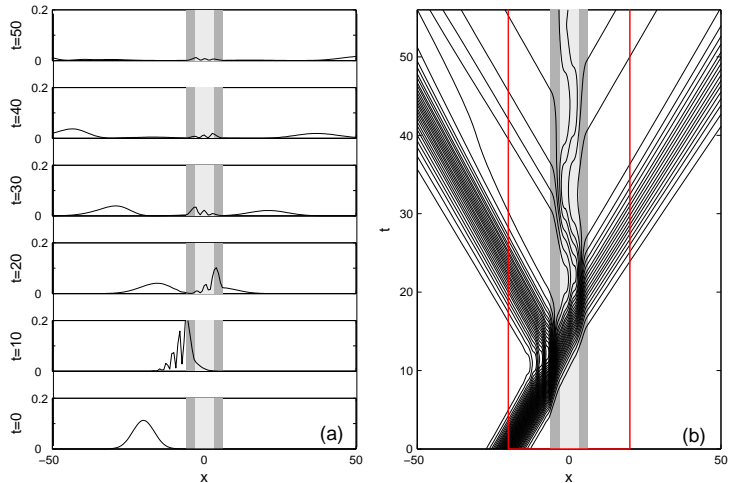}
\\
\caption{(a) Evolution of the probability density $\vert \Psi
\vert^2$. (b) Corresponding sample of 50 Bohmian trajectories. The
double potential barrier is indicated by the hatched
area.\label{DBtraj}}
\end{figure}

\begin{figure}[h!]
\centering
\includegraphics[width=.8\textwidth]{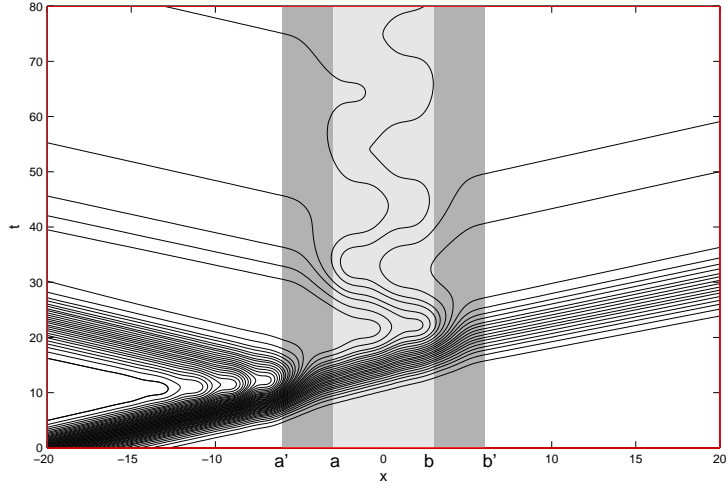}\\
\caption{Zoom into the region indicated in figure
\ref{DBtraj}(b).\label{DBtrajZoom}}
\end{figure}

As an example for a 1D scattering situation we consider the case
of a Gaussian wave packet impinging on a double potential barrier,
i.e. we are looking for solutions $\Psi$ to the Schr\"odinger
equation
\[
i \hbar \, \di_t \Psi= \left[ - \frac{\hbar^2}{2 m} \di_x^2 + V
\right] \Psi
\]
with
\[
V = V_0 \cdot \chi_{[a,b]}+ V_1 \cdot \left(
\chi_{[a',a[}+\chi_{]b,b']}\right).
\]
$\chi_{[\alpha,\beta]}$ denotes the characteristic function on the
interval $[\alpha,\beta] \in \R$ and $a'<a<b<b'$ (see figure
\ref{DBPotential}). The parameter reduction $\hbar,m,V_0 \to 1$,
e.g., is achieved by taking time in units of $\frac{\hbar}{2
V_0}$, space in units of $\frac{\hbar}{\sqrt{2 m V_0}}$. $V_1$ is
taken in units of $V_0$. In figure \ref{DBtraj}(a) the evolution
of the probability density $\vert \Psi \vert^2$ is given for the
case $a'=-6$, $a=-3$, $b=3$, $b'=6$ and $V_1=2$ in the chosen
units. The mean kinetic energy of the packet is $1.5^2 V_0=2.25
V_0$. In figure \ref{DBtraj}(b) a sample of 50 corresponding
Bohmian trajectories is illustrated. The initial distribution of
starting points of the trajectories resemble the initial Gaussian
distribution of the wave packet. Figure \ref{DBtrajZoom} shows a
zoom into the area indicated by the rectangle in figure
\ref{DBtraj}(b).

\noindent In figure \ref{DBtrajZoom} it becomes clear that, as
trajectories change their direction at $x=b$, the current density
in this case changes its sign also at the right edge of the area
in question. Therefore the restricted formulae of Oriols \emph{et
al} loose their validity and the generalized expressions
(\ref{TransTime}) and (\ref{ReflTime}) have to be applied.

\begin{figure}[h!]
\centering
\includegraphics[width=.8\textwidth]{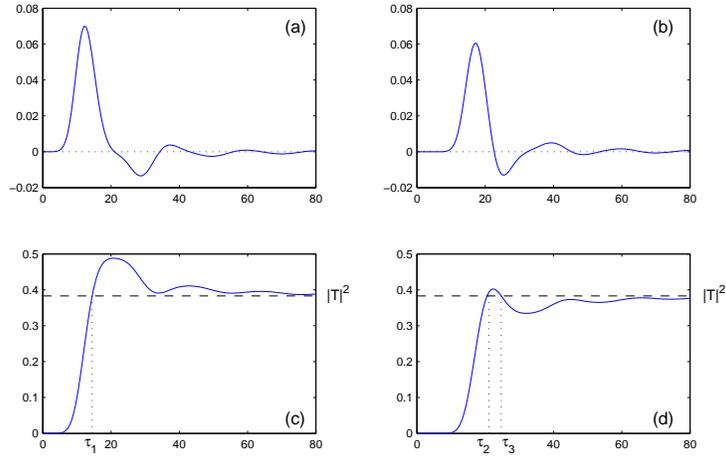}\\
\caption{Current densities $j(\cdot,a)$ (a) and $j(\cdot,b)$ (b)
and corresponding integrated current densities $f_a(\cdot)$ (c)
and $f_b(\cdot)$ (d).\label{DBjuf}}
\end{figure}

\noindent Figure \ref{DBjuf} shows the current densities and
corresponding integrated current densities at the edges $x=a$ and
$x=b$, respectively, as a function of time. In (c) and (d), the
transmission coefficient $\vert T \vert^2 =38.31\%$ is indicated
by the dashed line.

\begin{figure}[h!]
\centering
\includegraphics[width=.8\textwidth]{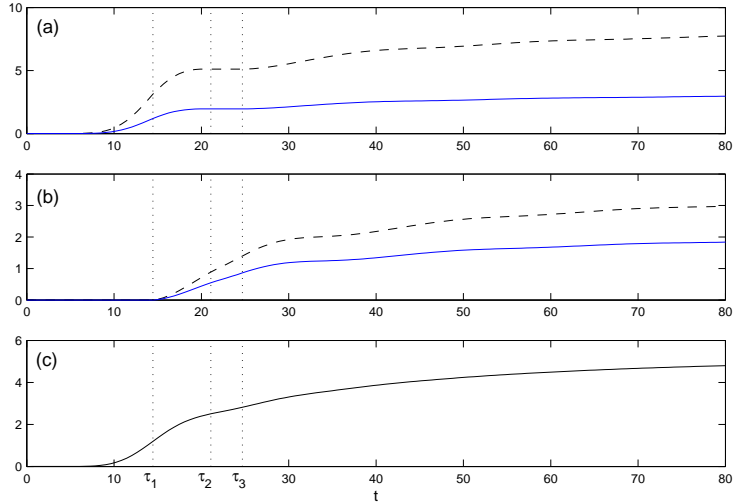}\\
\caption{(a) Transmission times $\mathcal{T}_T(t)$ (solid line)
and corresponding conditional transmission times (dashed line).
(b) Reflection times $\mathcal{T}_R(t)$ (solid line) and
conditional reflection times (dashed line). (c) Average dwell
times $\mathcal{T}_D(t)$ inside $[a,b]\times[0,t]$.
\label{DBDwellTimes}}
\end{figure}

Finally, in figure \ref{DBDwellTimes} the mappings
$\mathcal{T}_T(s)$ (a),  $\mathcal{T}_R(s)$ (b) and
$\mathcal{T}_D(s)$ (c) are illustrated, which give the Bohmian
transmission, reflection and overall average dwell times
$<\tau_X>$, $X \in \{T,R,D\}$ in $[a,b]$ from time $\ti=0$ onwards
as a function of the upper temporal bound $s=\tf$. In addition the
conditional transmission and reflection times, i.e. normalized to
the fraction of finally transmitted or reflected particles, are
indicated in (a) and (b).

Introducing as parameters the potential energy $V_0=0.25$ eV (i.e.
$V_1=0.5$ eV) and the effective electron mass $m_* = 0.07 m_e$
typical for $GaAs/GaAlAs$ double barrier heterostructures (see
e.g. \cite{HLref}), the temporal units get approximately $1.3$ fs,
the spatial units approximately $15$ $\AA$. Therefore the width of
the model heterostructure in figure \ref{DBPotential} is in the
range of $200$ $\AA$, which is easily achieved by the ultrathin
layers of modern semiconductor devices.

\noindent With the aid of formulae (\ref{TransTime}) and
(\ref{ReflTime}) of proposition \ref{propTRDT} the computational
effort involved with the calculation of transmission and
reflection times was reduced to about 5\% of that corresponding to
the computation by means of trajectory sampling.

\section*{Acknowledgements}

I'm much indebted to Gebhard Gr\"ubl for helpful discussions and
corrections and also to Peter Wagner for valuable suggestions
regarding the appendix, both University of Innsbruck. This work
has been supported by Doc-Fforte [Doctoral Scholarship Programme
of the Austrian Academy of Sciences].


\section*{Appendix A. Proof of equality (\ref{jrhoZshg})}\label{sProofjrhoZshg}

Recall that the scattering solutions under consideration are given
by $\Psi(t,\cdot)=\Omega^{in} \phi(t,\cdot)$, with
$\phi(t,\cdot)=e^{-i H_0 t/ \hbar} \phi_0$ a solution to the free
Schr\"odinger equation. The Fourier transform
$\varphi:=\mathcal{F}(\phi_0)$ is assumed to be localized
exclusively on the positive half line, for which reason
$\phi(t,\cdot)$ will be further and further localized to the left
for large negative times (in the sense of the $L^2$-norm $\Vert
\cdot \Vert$).

\begin{figure}[h!]
\centering
\includegraphics[width=.4\textwidth]{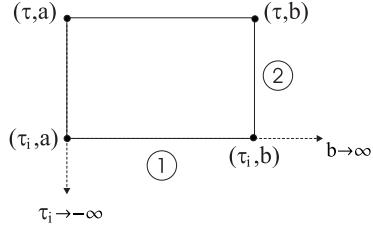}\\
\caption{Closed space time region.\label{F13proof}}
\end{figure}

Figure \ref{F13proof} then illustrates, that for the scattering
wave packet $\Psi$, formula (\ref{jrhoZshg}) is a plausible
conjecture. In the following a rigorous proof will be given for a
limited class of scattering solutions.

\noindent For the closed spacetime region in figure \ref{F13proof}
integration of the continuity equation assures that
\begin{eqnarray}
\int\limits_a^b \vert \Psi(\tau,\xi) \vert^2 \; d\xi & = &
\int\limits_{\tau_i}^{\tau} j(s,a) \; ds +
\underbrace{\int\limits_a^b \vert \Psi(\tau_i,\xi) \vert^2 \;
d\xi}_{\textcircled{1}} - \underbrace{\int\limits_{\tau_i}^{\tau}
j(s,b) \; ds}_{\textcircled{2}}. \nonumber
\end{eqnarray}
The aim is to show that the terms $\textcircled{1}$ and
$\textcircled{2}$ converge towards zero in the limits $\tau_i \to
-\infty$ and $b \to \infty$, in which case formula
(\ref{jrhoZshg}) applies.

\begin{prop}
For every scattering solution $\Psi$ with the above properties,
term \textcircled{1} vanishes in the limits $\tau_i \to -\infty$
and $b \to \infty$, i.e.
\begin{eqnarray}\label{rhointlim}
\lim_{\tau_i \to -\infty} \lim_{b \to \infty} \textcircled{1} & =
&  \lim_{M \to \infty} \int\limits_a^{M} \vert \Psi(-M,\xi)
\vert^2 \; d\xi = 0
\end{eqnarray}

\end{prop}

\begin{proof}
First note that the inequalities
\begin{eqnarray}\label{rhointlim2}
0 \leq \int\limits_a^{M} \vert \Psi(-M,\xi) \vert^2 \; d\xi \leq
\int\limits_a^{\infty} \vert \Psi(-M,\xi) \vert^2 \; d\xi
\end{eqnarray}
hold. For the given scattering solution with $\lim\limits_{t \to
-\infty} \left\Vert \Psi(t,\cdot) - \phi(t,\cdot) \right\Vert =
0$, according to Dollard \cite{Dollard}, the relation
\begin{eqnarray}
\lim_{t \to -\infty} \int\limits_A \left\vert \Psi(t,\xi)
\right\vert^2 \; d\xi = \lim_{t \to -\infty} \int\limits_A
\left\vert C_t(\phi_0)(\xi) \right\vert^2 \; d\xi \nonumber
\end{eqnarray}
holds for every measurable set $A \subseteq \R$ and with
\begin{eqnarray}
C_t(\phi_0)(x):=\left( \frac{m}{i \hbar t} \right)^{1/2} e^{i m
x^2/2 \hbar t} \varphi \left( \frac{mx}{\hbar t} \right).
\nonumber
\end{eqnarray}
With this, inequality (\ref{rhointlim2}) and $supp\{ \varphi \}
\subseteq \R_+$ one immediately proofs (\ref{rhointlim}) by
\begin{eqnarray}
\lim_{\tau_i \to -\infty} \int\limits_a^{\infty} \vert
\Psi(\tau_i,\xi) \vert^2 \; d\xi & = &  \lim_{\tau_i \to -\infty}
\int\limits_a^{\infty} \left\vert C_{\tau_i}(\phi_0)(\xi)
\right\vert^2 \; d\xi\nonumber\\
& = &  \lim_{\tau_i \to -\infty} \left\vert \frac{m}{\hbar \tau_i}
\right\vert \int\limits_a^{\infty} \left\vert \varphi \left(
\frac{m \xi}{\hbar \tau_i} \right) \right\vert^2 \; d\xi \nonumber
\\
& = & \lim_{\tau_i \to -\infty} \int\limits_{-\infty}^{\frac{m
a}{\hbar \tau_i}} \left\vert \varphi \left( k \right)
\right\vert^2 \; dk \nonumber \\
& = & \int\limits_{-\infty}^{0} \left\vert \varphi \left( k
\right) \right\vert^2 \; dk = 0. \nonumber
\end{eqnarray}

\end{proof}

\begin{prop}
For a freely evolving wave packet $\phi(t,\cdot)=e^{-i H_0 t/
\hbar} \phi_0$ with Fourier transform
$\varphi:=\mathcal{F}(\phi_0)$, $\varphi \in C^1(\R)$ and $supp\{
\varphi \} \subseteq [a_1,a_2]$, $0<a_1<a_2$ (i.e. $\varphi \in
C^1_0(\R_+)$) term \textcircled{2} vanishes in the limits $\tau_i
\to -\infty$ and $b \to \infty$, i.e.,

\begin{eqnarray}\label{jintlim}
\lim_{\tau_i \to -\infty} \lim_{b \to \infty} \textcircled{2}  & =
& \lim_{M \to \infty} \int\limits_{-M}^{\tau} j(s,M) \; ds = 0.
\end{eqnarray}
\end{prop}

\begin{proof}
Equation (\ref{jintlim}) can immediately be shown by a stationary
phase argument. In the following the parameter reduced notation
$\hbar=m=1$ will be used.

The solution $\phi$ of the free Schr\"odinger equation can be
written in the form
\begin{eqnarray}
\phi(t,x) & = & \F^{-1}\left( e^{-i \omega t} \F(\phi_0) \right)
(x)= (2 \pi)^{-1/2} \int dk \; e^{i (kx-\omega(k)t)} \varphi(k)
\nonumber
\end{eqnarray}
with the function $\omega:\R \to \R,\; k \mapsto k^2/2$ being in
$C^{\infty}(\R)$. The stationary phase argument then states (cf,
e.g., \cite{ReedSimon}, appendix 1 to XI.3) that for every open
set $A \supseteq [a_1,a_2] \supseteq \left\{ \omega'(k)\vert k \in
supp\{ \varphi \} \right\}$, there is a constant $C>0$ such that
for all $(t,x) \in \R^2$ with $\frac{x}{t} \notin A$
\[
\vert \phi(t,x) \vert \leq C \; (1 + \vert x \vert + \vert t
\vert)^{-1}.
\]
The same considerations then deliver a second constant $C'>0$ such
that for all $(t,x) \in \R^2$ with $\frac{x}{t} \notin A$
\[
\left\vert \frac{\di}{\di x} \phi(t,x)\right\vert = \left\vert
\int dk \; (i\,k) \cdot e^{i (kx-\omega(k)t)} \varphi(k)
\right\vert \leq C' \; (1 + \vert x \vert + \vert t \vert)^{-1}.
\]
We choose without loss of generality
$A:=\left]\frac{a_1}{2},2\cdot a_2\right[$. Then for a fixed $\tau
\in \R$ there is a $M>0$ such that for all $t \leq \tau$ and all
$x \geq M$: $\frac{x}{t} \notin A$ (set, e.g., $M \geq 2  a_2
\tau$). Then, $\forall x \geq M$ and $\forall t\leq \tau$
\begin{eqnarray}
\vert j(t,x) \vert & \leq & \left\vert \phi(t,x) \right\vert \cdot
\left\vert \frac{\di}{\di x} \phi(t,x) \right\vert \; \leq \, C
\cdot C' \cdot (1 + \vert x \vert + \vert t \vert)^{-2}. \nonumber
\end{eqnarray}
Therefore
\begin{align}
&\lim_{M \to \infty} \left\vert \int_{-M}^{\tau} j(t,M) \; dt
\right\vert \leq  \lim_{M \to
\infty} \int_{-M}^{\tau} \vert j(t,M) \vert \; dt \nonumber \\
&\leq C \cdot C' \; \lim_{M \to \infty} \int_{-M}^{\tau} (1+
\vert M \vert + \vert t \vert)^{-2} \; dt \nonumber \\
&= C \cdot C' \; \lim_{M \to \infty} \left(
\frac{1+sgn(\tau)}{1+\vert M \vert} - \frac{1}{1+2 \vert M \vert}
- \frac{sgn(\tau)}{1+ \vert M \vert +\vert \tau \vert} \right) =
0. \nonumber
\end{align}
\end{proof}

\noindent Now consider scattering solutions
\[
\Psi(t,x)=\int\limits_{0}^{\infty} dk \; \varphi(k) \,
\tilde{\phi}^{in}(k;x) \, e^{-i k^2 t/2}
\]
with $\tilde{\phi}^{in}(k;x)$ a solution
to the corresponding Lippman-Schwinger equation and $\varphi$
again exclusively localized on the positive half-line. If the
$\tilde{\phi}^{in}(k;x)$ have the form
\begin{equation}\label{LSseparation}
\tilde{\phi}^{in}(k;x) = T(k) \, e^{i k x}
\end{equation}
for $x>R$ for some $R>0$, and $T \in C^1(\R)$ (e.g., the potential
barrier), then the above line of reasoning applies analogously.
Clearly expression (\ref{LSseparation}) is only valid for
scattering potentials with support bounded from the right.

\section*{Appendix B. Proof of proposition \ref{propTRDT}}\label{sProofpropTRDT}

By $\gamma_x: \R \rightarrow \R, \; t \mapsto \gamma_x(t)$, the
integral curve of the Bohmian velocity vector field with initial
datum $x$ is denoted. The intervals $[a_{t},b_{t}]:=\{ x \in \R /
\gamma_x(t) \in [a,b]\} $ represent the initial data on
configuration space, which are projected onto the interval $[a,b]$
at time $t$ along their integral curves. Let $x_c \in \R$ be the
initial condition of the trajectory, which separates the
transmitted from the reflected ensemble. Then the transmission
time (\ref{transT}) gets

\begin{eqnarray}
\langle \tau_T \rangle & = & \int\limits_{x_c}^{\infty} d\xi \;
\vert \Psi(0,\xi) \vert^2
\int\limits_{\ti}^{\tf} dt \; \chi_{[a,b]}(\gamma_{\xi}(t)) \nonumber \\
 & = & \int\limits_{\ti}^{\tf} dt \int\limits_{x_c}^{\infty} d\xi \; \vert \Psi(0,\xi)
\vert^2 \; \chi_{[a_{t},b_{t}]}(\xi) \nonumber \\
& = & \int\limits_{[\ti,\tf]} \!\! dt \!\!\!
\int\limits_{[a_{t},b_{t}] \cap [x_c,\infty[}
\!\!\!\!\!\!\!\!\!\!\! d\xi \; \vert \Psi(0,\xi) \vert^2
\nonumber \\
& = & \int\limits_{[\ti,\tf]} dt  \int\limits_{[a,b]} d\xi \;
\vert \Psi(t,\xi) \vert^2 \; \Theta(\xi-\gamma_{x_c}(t)).
\nonumber
\end{eqnarray}
Analogously, the reflection time reads as
\begin{eqnarray}
\langle \tau_R \rangle &  = & \int\limits_{[\ti,\tf]} dt
\int\limits_{[a,b]} d\xi \; \vert \Psi(t,\xi) \vert^2 \;
\Theta(\gamma_{x_c}(t)-\xi). \nonumber
\end{eqnarray}
Now the right hand side of (\ref{TransTime}) together with
(\ref{xc}) and (\ref{jrhoZshg}) gets
\begin{eqnarray*}
&& \int_{\ti}^{\tf} dt \; \left[ \min\left\{ \int_{a}^{\infty}
\vert \Psi(t,\xi) \vert^2 \; d\xi, \int_{\gamma_{x_c}(t)}^{\infty}
\vert \Psi(t,\xi) \vert^2 \; d\xi \right\} \right. \\
&&- \left. \min\left\{ \int_{b}^{\infty} \vert \Psi(t,\xi) \vert^2
\; d\xi, \int_{\gamma_{x_c}(t)}^{\infty}
\vert \Psi(t,\xi) \vert^2 \; d\xi \right\} \right] \\
&=& \int_{\ti}^{\tf} dt \; \left[
\int_{\max\{a,\gamma_{x_c}(t)\}}^{\infty} \vert \Psi(t,\xi)
\vert^2 \; d\xi - \int_{\max\{b,\gamma_{x_c}(t)\}}^{\infty}
\vert \Psi(t,\xi) \vert^2 \; d\xi \right] \\
&=& \int_{\ti}^{\tf} dt
\int_{\max\{a,\gamma_{x_c}(t)\}}^{\max\{b,\gamma_{x_c}(t)\}}
\vert \Psi(t,\xi) \vert^2 \; d\xi\\
&=& \int_{\ti}^{\tf} dt \int_{a}^{b} \vert \Psi(t,\xi) \vert^2
\cdot \Theta(\xi-\gamma_{x_c}(t)) \; d\xi = <\tau_T>.
\end{eqnarray*}
Analogously the right hand side of (\ref{ReflTime}) together with
(\ref{xc}) and (\ref{jrhoZshg}) becomes
\begin{eqnarray*}
&&\int_{\ti}^{\tf} dt \; \left[ \max\left\{ \int_{a}^{\infty}
\vert \Psi(t,\xi) \vert^2 \; d\xi, \int_{\gamma_{x_c}(t)}^{\infty}
\vert \Psi(t,\xi)
\vert^2 \; d\xi \right\} \right. \\
&&\left.-  \max\left\{ \int_{b}^{\infty} \vert \Psi(t,\xi) \vert^2
\; d\xi, \int_{\gamma_{x_c}(t)}^{\infty}
\vert \Psi(t,\xi) \vert^2 \; d\xi \right\} \right] \\
&=& \int_{\ti}^{\tf} dt
\int_{\min\{a,\gamma_{x_c}(t)\}}^{\min\{b,\gamma_{x_c}(t)\}}
\vert \Psi(t,\xi) \vert^2 \; d\xi\\
&=& \int_{\ti}^{\tf} dt \int_{a}^{b} \vert \Psi(t,\xi) \vert^2
\cdot \Theta(\gamma_{x_c}(t)-\xi) \; d\xi = <\tau_R>
\end{eqnarray*}
which completes the proof. \hfill \rule{0.5em}{0.5em}



\end{document}